\documentclass[aps,prl,reprint,amsmath,amssymb,nofootinbib,floatfix,superscriptaddress]{revtex4-2}
\usepackage{graphicx}
\usepackage{dcolumn}
\usepackage{bm}
\usepackage{color}
\usepackage{tabularx}
\usepackage{booktabs}
\usepackage{orcidlink}
\usepackage{hyperref}
\hypersetup{colorlinks=true,citecolor= blue}

\usepackage{enumerate}
\usepackage{enumitem}

\newcommand{\beginsupplement}{
  \setcounter{table}{0}
  \renewcommand{\thetable}{S\arabic{table}}%
  \setcounter{figure}{0}
  \renewcommand{\thefigure}{S\arabic{figure}}%
  \setcounter{equation}{0}
  \renewcommand{\theequation}{S\arabic{equation}}%
}

\def\Xint#1{\mathchoice
       {\XXint\displaystyle\textstyle{#1}}%
       {\XXint\textstyle\scriptstyle{#1}}%
       {\XXint\scriptstyle\scriptscriptstyle{#1}}%
       {\XXint\scriptscriptstyle\scriptscriptstyle{#1}}%
       \!\int}
    \def\XXint#1#2#3{{\setbox0=\hbox{$#1{#2#3}{\int}$}
         \vcenter{\hbox{$#2#3$}}\kern-.5\wd0}}
    
    \def\dashint{\Xint-}
    
\begin{document}

\title{Can the strong interactions between hadrons be determined using femtoscopy?}

\author{Evgeny Epelbaum\orcidlink{0000-0002-7613-0210}}
\email{evgeny.epelbaum@ruhr-uni-bochum.de}
\affiliation{Ruhr-Universit\"at Bochum, Fakult\"at f\"ur Physik und Astronomie, Institut f\"ur Theoretische Physik II, D-44780 Bochum, Germany}

\author{Sven Heihoff\orcidlink{0009-0000-3641-0640}}
\email{sven.heihoff@ruhr-uni-bochum.de}
\affiliation{Ruhr-Universit\"at Bochum, Fakult\"at f\"ur Physik und Astronomie, Institut f\"ur Theoretische Physik II, D-44780 Bochum, Germany}

\author{Ulf-G. Mei{\ss}ner\orcidlink{0000-0003-1254-442X}}
\email{meissner@hiskp.uni-bonn.de}
\affiliation{Helmholtz-Institut f\"ur Strahlen- und Kernphysik,
  Rheinische Friedrich-Wilhelms Universit\"at Bonn, D-53115 Bonn, Germany}
\affiliation{Bethe Center for Theoretical Physics, Rheinische Friedrich-Wilhelms Universit\"at Bonn, D-53115 Bonn, Germany}
\affiliation{Center for Science and Thought, Rheinische Friedrich-Wilhelms Universit\"aat Bonn, D-53115 Bonn, Germany}
\affiliation{Institute for Advanced Simulation (IAS-4), Forschungszentrum J\"ulich, D-52425 J\"ulich, Germany}
\affiliation{Peng Huanwu Collaborative Center for Research and Education,
International Institute for Interdisciplinary and Frontiers, Beihang University, Beijing 100191, China}

\author{Alexander Tscherwon}
\email{alexander.tscherwon@ruhr-uni-bochum.de}
\affiliation{Ruhr-Universit\"at Bochum, Fakult\"at f\"ur Physik und Astronomie, Institut f\"ur Theoretische Physik II, D-44780 Bochum, Germany}

\begin{abstract}
In the last decades, femtoscopic measurements from heavy-ion collisions have become a popular tool to
investigate the strong interactions between hadrons. The key observables measured in such experiments
are the two-hadron momentum correlations, which depend on the production mechanism of hadron pairs
and the final-state interactions. Given the complexity of ultra-relativistic collision experiments,
the source term describing the production mechanism can only be modeled phenomenologically
based on numerous assumptions. The commonly employed approach for analyzing femtoscopic data
relies on the Koonin-Pratt formula, which relates the measured correlation functions with
the relative wave function of an outgoing hadron pair and a source term that
is assumed to be universal.
Here, we critically examine this universality assumption and show that for strongly interacting
particles such as nucleons,
the interpretation of femtoscopic measurements suffers from a potentially large intrinsic uncertainty.  
We also comment on the ongoing efforts to explore three-body interactions using this experimental technique. 
\end{abstract}

\maketitle
{\it Introduction.---}Femtoscopic measurements are becoming increasingly popular as a tool to study
hadronic interactions and constitute an integral part of the ongoing and upcoming experimental
programs at high-energy heavy-ion facilities such as RHIC, LHC, GSI/FAIR and J-PARC-HI. Originally
developed as an imaging technique to extract spatial and temporal characteristics of particle
production mechanisms in ultra-relativistic heavy-ion collisions \cite{Koonin:1977fh,Bauer:1992ffu,Pratt:1998vr,Lisa:2005dd},
femtoscopy has nowadays evolved into a method for measuring the strong interactions between hadrons
\cite{Fabbietti:2020bfg,ALICE:2020mfd,STAR:2018uho,STAR:2024zvj,Stefaniak:2024jcr}. A traditional way of exploring hadron structure and dynamics
by means of scattering experiments aims at the determination of the corresponding on-shell scattering
amplitudes, which are unambiguously defined and experimentally measurable quantities. Unfortunately,
the short lifetime of hadronic resonances makes it difficult to perform scattering experiments beyond
the lightest hadrons. For example, there is a wealth of experimental data on nucleon-nucleon (NN)
scattering, while only a handful of/no data are available for hyperon-nucleon/hyperon-hyperon
scattering. Femtoscopic measurements from high-energy proton-proton (pp) or heavy-ion collisions
do not require preparing a beam of unstable particles and thus provide an attractive alternative
to scattering experiments. This, however, comes at the price of dealing with a complicated
initial state after hadronization, whose modeling becomes an integral part of the analysis
of the experimental data. After making several approximations and assumptions \cite{Lisa:2005dd},
the two-particle correlation function $C({\bf k} )$ in the center-of-mass system (cms)
is usually written in the form \cite{Koonin:1977fh,Lednicky:1981su,Lisa:2005dd}
\begin{equation}
\label{KooninPratt}
C ({\bf k}  ) = \int d {\bf r}  \, S_{12} ({\bf r}  ) \, \big| \Psi ({\bf r} , \, {\bf k}  ) \big|^2\,,
\end{equation}
where ${\bf k} $ is the relative momentum, $S_{12} ({\bf r}  )$ is the source function and
$\Psi ({\bf r} , \, {\bf k}  )$ is the relative wave function of the outgoing two-body state,
which coincides with the stationary solution of the scattering problem $\Psi ({\bf r}, \, {\bf k} )
= \langle {\bf r} | \Psi^{(+)}_{- \bf k} \rangle$ normalized to yield $\langle {\bf r} | \Psi^{(+)}_{- \bf k}
\rangle \to e^{- i {\bf k} \cdot {\bf r}}$ in the absence of the interaction, see e.g.~Ref.~\cite{Lednicky:2005tb}.
This is the celebrated Koonin-Pratt formula, which is widely utilized to analyze femtoscopic measurements.
The pathway from experiment to interpretation can then be schematically summarized as follows
(see Refs.~\cite{Fabbietti:2020bfg,ALICE:2022wpn} for details):  
\begin{itemize}
\vspace{-0.13cm}  
\item[i.] Measurement of the correlation functions $C({\bf k}  )$. \\[-16pt]
\item[ii.] Modeling of the source function $S_{12} ({\bf r}  )$ which is deemed to be universal.
Here, one usually assumes a spherically symmetric Gaussian form, whose size $r_0$ is extracted from
experimental data on $C ({\bf k} )$ using some model to describe the strong interaction. In
particular, $r_0$ is often fitted to pp correlation functions using the Reid Soft-Core \cite{Reid:1968sq}
or Argonne $v_{18}$ \cite{Wiringa:1994wb} potential models \cite{Mihaylov:2018rva}. A more
sophisticated modeling of the source is described in Ref.~\cite{ALICE:2020ibs}. \\[-16pt]
\item[iii.] Once the source function $S_{12} ({\bf r}  )$ is fixed, Eq.~(\ref{KooninPratt})
allows one to probe hadronic interaction models using experimental data on $C ({\bf k}  )$.  
\end{itemize}
\vspace{-0.13cm}  
It goes beyond the scope of this paper to address the validity of various assumptions and
approximations used to arrive at the formula~(\ref{KooninPratt}), which are, in fact, well documented
in the literature \cite{Pratt:1997pw,Anchishkin:1997tb,Lisa:2005dd,Lednicky:2005tb}. Rather, we would
like to point out that the way Eq.~\eqref{KooninPratt} is used to probe hadronic interactions as
outlined above suffers from a fundamental flaw: Combined with the universality assumption for
the source function $S_{12} ({\bf r}  )$, it implies the measurability of hadronic wave functions
and thus also of the corresponding interaction potentials. Yet, hadronic potentials are well known
{\it not} to represent observable quantities, see e.g.~Ref.~\cite{Birse:2012ph} for a  recent
discussion. While interaction potentials can, at least in principle, be determined {\it ab initio}
using lattice QCD \cite{Aoki:2012tk}, their form depends on the choice of interpolating fields
for parametrizing hadrons in terms of valence quarks. This inherent scheme dependence does,
of course, not affect observable quantities like S-matrix, as can be shown by virtue of the
LSZ reduction formalism \cite{Lehmann:1954rq}. 
 
To make the essence of the problem more explicit while keeping the presentation simple, we
restrict ourselves to non-relativistic systems in the framework of quantum mechanics and
start with rewriting Eq.~(\ref{KooninPratt}) in a more general form
\begin{equation}
\label{KooninPrattGeneralized}
C ({\bf k}  ) =  \langle \Psi_{{- \bf k} }^{(+)} | \hat S_{12}  | \Psi_{{- \bf k} }^{(+)} \rangle\,.
\end{equation}
The Koonin-Pratt formula~(\ref{KooninPratt}) is the coordinate-space representation of
Eq.~\eqref{KooninPrattGeneralized}, subject to the restriction that the source term is local,
$\langle {\bf r}' | \hat S_{12} | {\bf r} \rangle = \delta ({\bf r}' - {\bf r}) S_{12} ({\bf r})$.
Eq.~\eqref{KooninPrattGeneralized} makes it explicit that the correlation functions are invariant
under a change of basis in the Hilbert space induced by unitary transformations (UTs) $\hat U$
as required for any observable quantity\footnote{Similarly to gauge or coordinate transformations, such unitary
transformations change the interactions and thus also (unobservable)
off-shell quantities, while not affecting the measurable on-shell S-matrix.}:  
\begin{eqnarray}
  \label{KooninPrattUT}
  C ({\bf k}  ) &=&  \big( \langle \Psi_{{- \bf k} }^{(+)} | \hat U^\dagger \big) \big( \hat U \hat S_{12}
  \hat U^\dagger \big) \big( \hat U | \Psi_{{- \bf k} }^{(+)} \rangle \big) \nonumber \\
  &\equiv&  \langle \Psi_{{- \bf k} }'^{(+)} | \hat S_{12}'  | \Psi_{{- \bf k} }'^{(+)} \rangle\,.
\end{eqnarray}
However, it also shows that the source term cannot be assumed to be universal across interaction models,
and its form must be off-shell consistent with that of the interaction under consideration. Failure
to maintain off-shell consistency between the source and the wave function (or interaction
potential) by assuming $\hat S_{12}' = \hat S_{12}$ leads to model dependence of the calculated correlation
functions.  Given the quest for precision in femtoscopy experiments~\cite{ALICE:2020mfd}, it is important to
quantify model dependence in $C ({\bf k})$ caused by the assumed universality of the source.  

{\it Gedankenexperiment.---}For demonstration purposes, we focus here on two stable distinguishable
spinless particles interacting via a short-range force (but all arguments made here remain valid
for identical particles, nonvanishing spin and/or if the Coulomb
interaction is included,
  see the Supplemental Material~\cite{SuppMat} for
  a generalization to the proton-proton system)
  and
perform a Gedankenexperiment by letting Alice and Bob analyze two-body correlations. 

\begin{figure}[t]
    \centering
    \includegraphics[width=0.99\linewidth]{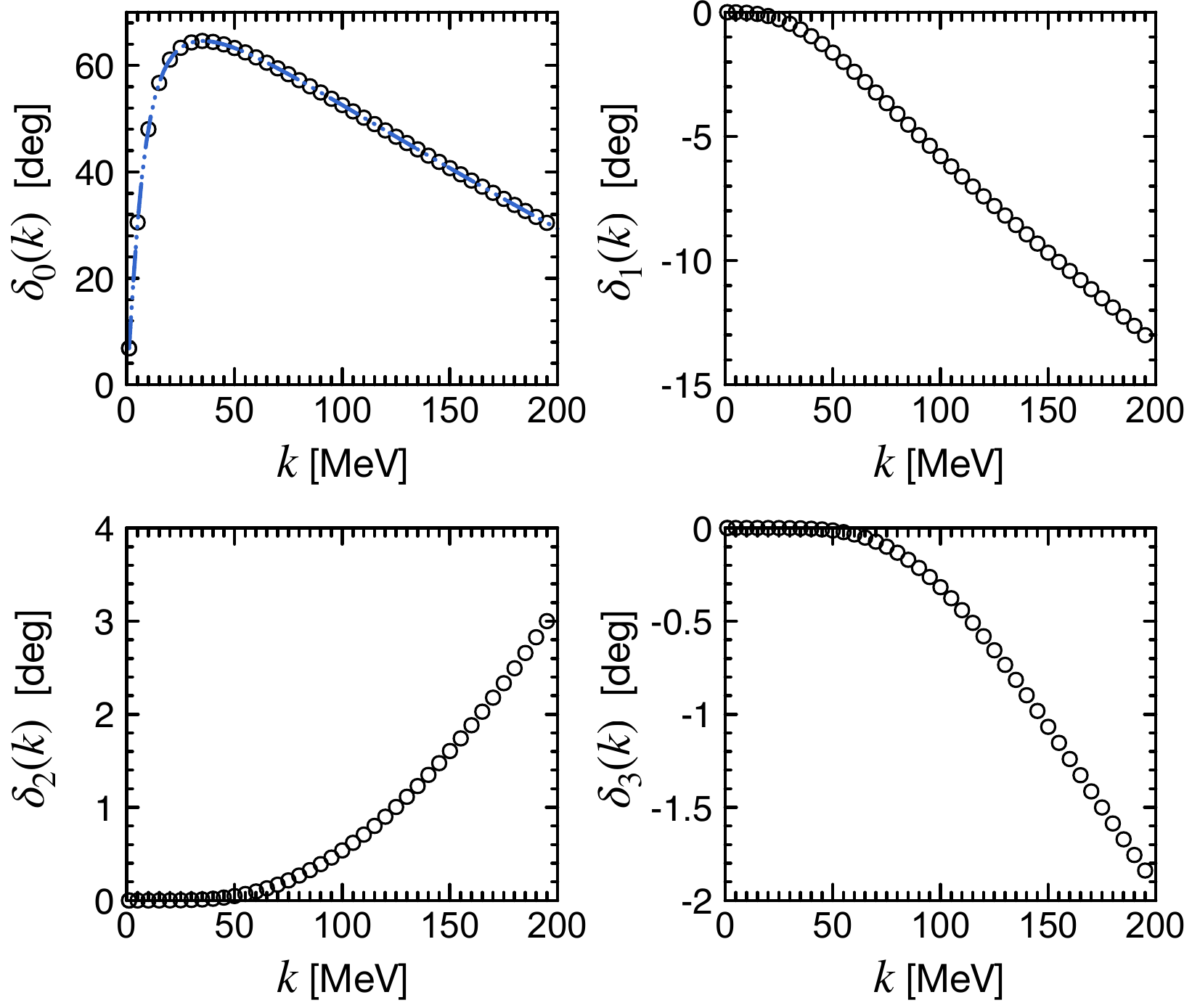}
    \caption{(Color online). S-, P-, D- and F-wave phase shifts for the interaction $V_{\rm Alice}$
      are shown with open circles as functions of the cms momentum.  The overlapping blue dotted and
      dashed lines show the phase shifts obtained from $V_{\rm Bob\mbox{-}I}$ and $V_{\rm Bob\mbox{-}II}$, respectively.
    }
    \label{fig0}
\end{figure}

\begin{figure}[t]
    \centering
    \includegraphics[width=0.99\linewidth]{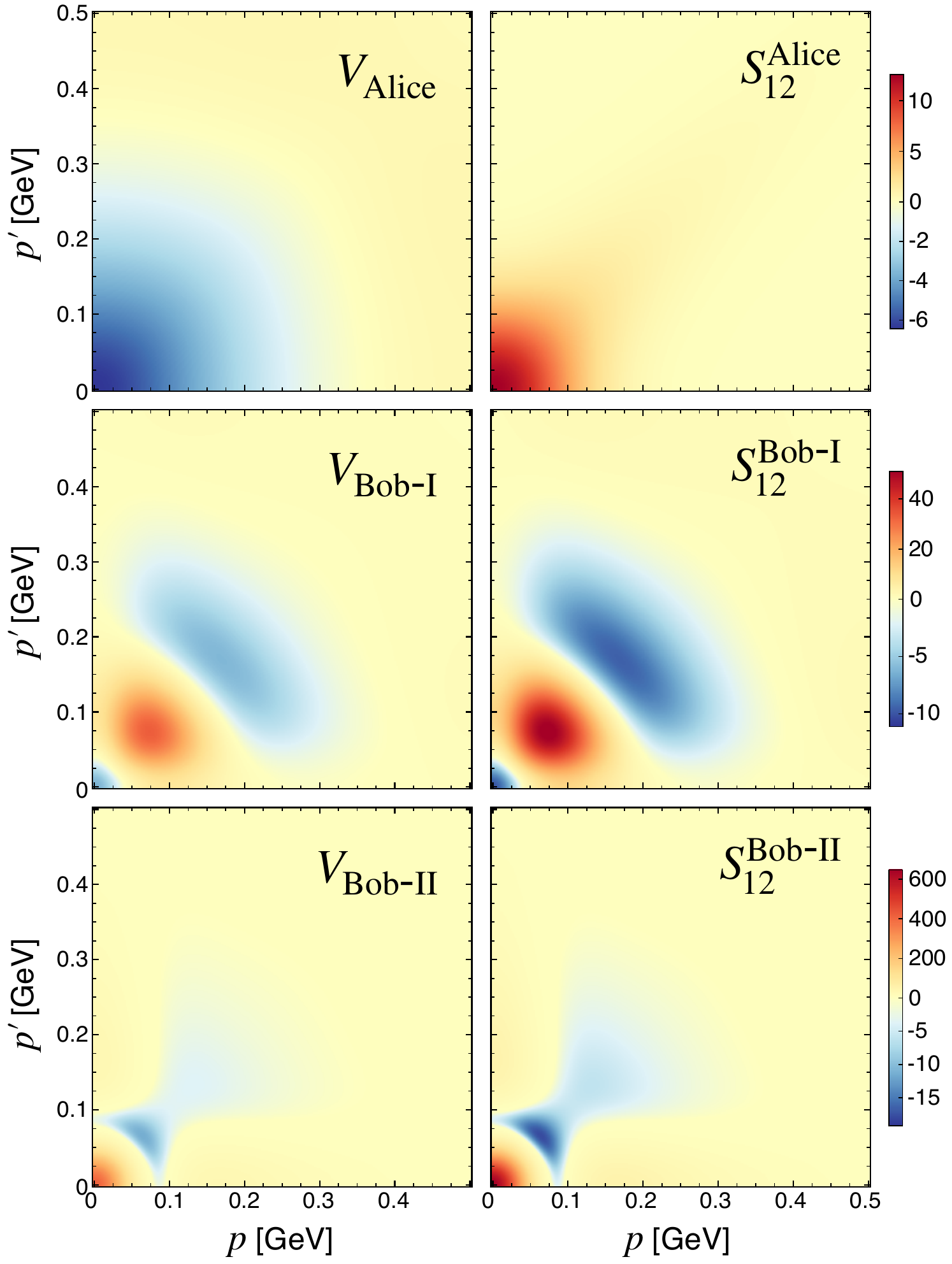}
    \caption{(Color online). S-wave momentum-space matrix elements of the potential in units of
      GeV$^{-2}$ (left column) and of the dimensionless source term (right column) as chosen
      by Alice (upper row) and seen by Bob (middle and bottom rows).
    }
    \label{fig1}
\end{figure}

The interaction between particles considered by Alice coincides with the spin-$0$ projection
of the neutron-proton chiral effective field theory (EFT) potential at fifth order (N$^4$LO$^+$)
\cite{Reinert:2017usi} with the cutoff parameter $\Lambda = 450$~MeV. That is,
\begin{equation}
\label{InteractionAlice}  
\langle p' l | \hat V_{\rm Alice} | p l \, \rangle \; =\;  \langle {p'}  l  | \hat V_{\rm np, \; N^4LO^+}^{s=0}
| p \,  l\rangle,
\end{equation}
where $p \equiv |{\bf p}|$ and $p' \equiv |{\bf p}'|$ with ${\bf p}$ and ${\bf p}'$ the initial and
final momenta in the cms, respectively,  while $l$ is the orbital angular momentum. The corresponding
phase shifts in the $l \leq 3$ partial waves are shown by black open circles in Fig.~\ref{fig0}.
Note that we use natural units with $\hbar = c = 1$ throughout this paper. 
For the source term, Alice adopts the commonly chosen Gaussian form  
\begin{equation}
\label{SourceAlice}
S_{12}^{\rm Alice} ({\bf r})\; =\;  \frac{e^{-r^2/(4 r_0^2)} }{(4 \pi r_0^2)^{3/2}} \;\; \Rightarrow \;
\langle {\bf p}' |\hat S_{12}^{\rm Alice} | {\bf p} \rangle = e^{-q^2 r_0^2} \,,
\end{equation}
where $q \equiv |{\bf q}|$ with ${\bf q} = {\bf p}' - {\bf p}$ the momentum transfer. The radius $r_0$
is set to $r_0 = 1.5$~fm.
The S-wave momentum-space matrix elements of the potential and the source term used by Alice and
specified in Eqs.~\eqref{InteractionAlice} and \eqref{SourceAlice} are visualized in the upper
row of Fig.~\ref{fig1}. 

\begin{figure}[t]
    \centering
    \includegraphics[width=0.99\linewidth]{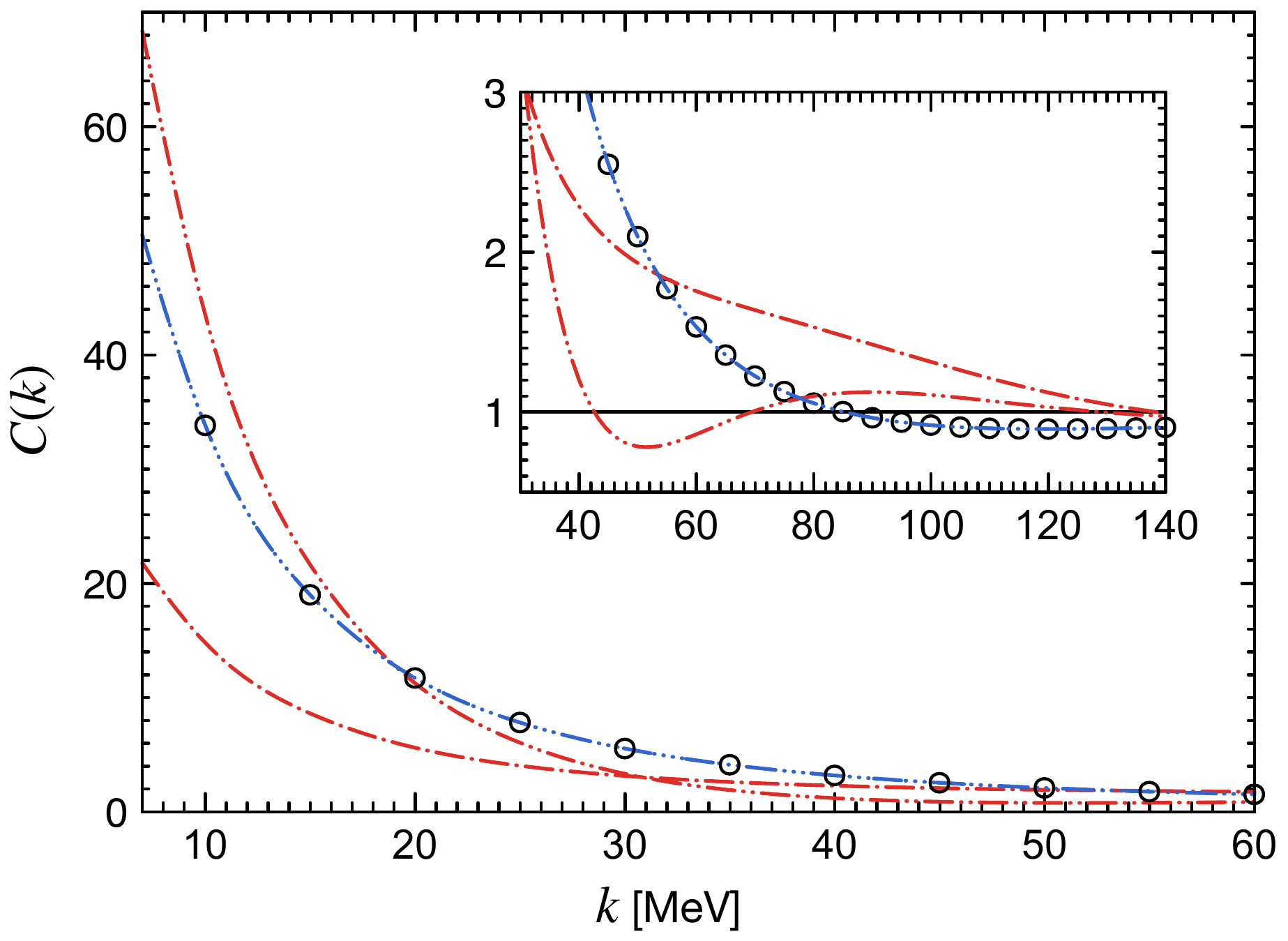}
    \caption{(Color online). Correlation functions calculated by Alice and Bob. The results of
      Alice are shown by open circles. Red dash-dotted and dash-double-dotted lines show the
      correlation functions calculated by Bob using the interactions  $V_{\rm Bob\mbox{-}I}$ and
      $V_{\rm Bob\mbox{-}II}$, respectively, and the source term $S_{12}^{\rm Alice}$ (assumed to be universal).
      The overlapping blue dotted and dashed lines show the results of Bob using the
      properly transformed source terms $S_{12}^{\rm Bob\mbox{-}I}$ and $S_{12}^{\rm Bob\mbox{-}II}$, respectively. 
    }
    \label{fig2}
\end{figure}

The correlation functions are usually calculated in coordinate space. Under the commonly made
assumption that particles interact only in the $l=0$ state, Eq.~\eqref{KooninPratt} can,
for a spherically symmetric source function, be reduced to (see, e.g., Ref.~\cite{ExHIC:2017smd}):
\begin{equation}
  \label{CkApprox}
C (k) = 1 + \int_0^\infty 4 \pi r^2 dr S_{12}(r) \Big[ |\Psi (r, \, k) |^2 - | j_0 (kr)|^2 \Big],
\end{equation}  
with $j_0 (kr)$ the spherical Bessel function and the S-wave scattering wave function $\Psi (r, \, k)$
is normalized according to
\begin{equation}
\Psi (r, \, k) \stackrel{r \to \infty}{\longrightarrow} \frac{1}{2 i k r} \Big( - e^{- 2 i \delta_0 (k)}
e^{- i k r} + e^{i k r} \Big).
\end{equation}
Here, $\delta_0 (k)$ is the S-wave phase shift.
The wave-function  $\Psi (r, \, k)$ can be obtained from the half-shell T-matrix,
see e.g.~Ref.~\cite{Haidenbauer:2021zvr}. 
However, to avoid numerical integrations of strongly oscillating
functions, it is more convenient to directly compute the matrix element in Eq.~\eqref{KooninPrattGeneralized}
in the partial-wave momentum-space basis. For a spherically symmetric
source, we find (see also Ref.~\cite{Albaladejo:2024lam}):
\begin{eqnarray}
\label{CkMomSpace}
C(k ) &=& \sum_l \frac{2l +1}{4 \pi} \, \cos^2 \delta_l (k) \, \Big[S_{kk}^l + K_{kp}^l  \circ S_{pk}^l
\nonumber \\
     &+&  S_{kp}^l  \circ K_{pk}^l  + K_{kp}^l  \circ S_{pp'}^l \circ  K_{p'k}^l \Big],
\end{eqnarray}
where we have introduced a short-hand notation $S_{pp'}^l \equiv \langle p l| \hat S_{12}| p' l \rangle$,
$K_{pk}^l = K_{kp}^l\equiv \langle p l| \hat K| k l \rangle$ and
\begin{eqnarray}
A_{pp'}^l  \circ B_{p'p''}^l  & \equiv & \dashint_0^\infty p'^2 dp' \, A_{pp'}^l  \, \frac{2\mu}{k^2 - p'^2}
\,  B_{p'p''}^l\,.
\end{eqnarray}
Here, $\mu$ is the reduced mass and $\dashint$ denotes the Cauchy principal value integral.
The half-shell K-matrix can be easily calculated for arbitrary short-range interactions by
solving the Lippmann-Schwinger integral equation 
\begin{eqnarray}
\label{LSeqnKmat}
K_{pk}^l &=& V_{pk}^l + V_{pp'}^l \circ K_{p'k}^l
\end{eqnarray}
using standard numerical methods. The on-shell K-matrix is related to
the S-matrix via
$S_l(k) = e^{2 i \delta_l (k)}  = (1 - i \mu \pi k K^l_{kk})/ (1+ i
\mu \pi k K^l_{kk})$.
We note in passing that Eq.~\eqref{CkMomSpace} can be brought to the form of Eq.~\eqref{CkApprox}
if the source term is local and the interaction vanishes in all $l \neq 0$ channels.  
The correlation function calculated by Alice using Eq.~\eqref{CkMomSpace} is shown by open circles
in Fig.~\ref{fig2}.
We have also checked these results by  using Eq.~\eqref{KooninPratt} directly. 

On the other hand, Bob performs his analysis using a different basis in the Hilbert space with
$|\Psi_{\rm Bob} \rangle = \hat U | \Psi_{\rm Alice} \rangle$, where the unitary operator is taken
as a rank-1 separable form
\begin{eqnarray}
\label{SauerUnitarytrafo}  
\hat U &=& 1 -2 | g \rangle \langle g |, \quad  \langle g | g \rangle = 1\,,
\end{eqnarray} 
and $|g\rangle$ is chosen following Refs.~\cite{Haftel:1971er,Sauer:1974jh} as
\begin{eqnarray}
g ({\bf r}) &\equiv& \langle {\bf r}| g\rangle = C r (1 - \beta r)e^{- \alpha r} \,,
\end{eqnarray}
with $C$ the normalization constant fixed from $\int d{\bf r} \big[g ({\bf r})\big]^2 = 1$.
The parameter $\alpha$ specifies the inverse range of the transformation, while $\beta^{-1}$
gives the position of the node in the
profile function $g(r)$. The interaction $V_{\rm Alice}$ in Bob's convention has the form 
\begin{eqnarray}
\label{UT_potential}
\hat V_{\rm Bob} &=& \hat U \bigg( \frac{\hat p^2}{2\mu} + \hat V_{\rm Alice} \bigg) \hat U^\dagger
- \frac{\hat p^2}{2\mu} \,,
\end{eqnarray}
where $\hat {\bf p}$ is the cms momentum operator. It can  easily be calculated numerically
in momentum space. Notice that the considered transformation acts only in the S-wave. 
To be specific, consider two different transformations corresponding to the parameters
$\alpha = 1.0\mbox{ fm}^{-1}$, $\beta = 0.25\mbox{ fm}^{-1}$ (set-I) and  $\alpha = 0.7\mbox{ fm}^{-1}$,
$\beta = 2.0\mbox{ fm}^{-1}$ (set-II). 
These transformations significantly affect the S-wave momentum-space matrix elements of the
interaction as shown in Fig.~\ref{fig1}. Yet, physical observables are, of course, independent of
the choice of basis in the Hilbert space. In particular, we have verified that the phase shifts
calculated from $\hat V_{\rm Alice}$, 
$\hat V_{\rm Bob\mbox{-}I}$ and $\hat V_{\rm Bob\mbox{-}II}$ agree with each other to machine precision,
see Fig.~\ref{fig0}.
In contrast, the correlation functions calculated by Bob under the assumption of the universal
source term, $\hat S^{\rm Bob}_{12} = \hat S^{\rm Alice}_{12}$, are as expected
different from the result found by Alice, as depicted by the red lines in Fig.~\ref{fig2}.
To restore the result for $C(k)$ obtained by Alice, the source term
needs to be brought into Bob's convention by applying the UT, $\hat S^{\rm Bob}_{12} = \hat U
\hat S^{\rm Alice}_{12}  \hat U^\dagger$, see the right column of Fig.~\ref{fig1} and blue lines in Fig.~\ref{fig2}.

{\it Scheme dependence in chiral EFT.---}The above example shows that off-shell ambiguities of
the interaction potentials can result in a large model dependence of the correlation function
if the source term is assumed to be universal. On the other hand,
realistic models of had\-ro\-nic interactions are constrained by physical principles like,
e.g., pion-exchange dominance at large distances and often share similarities when it comes to
modeling of the short-distance behavior. The remaining off-shell ambiguities may therefore be
expected to be less pronounced in practice.
In this context, it is instructive to examine how scheme dependence manifests itself in
chiral EFT for nuclear forces,
\cite{Epelbaum:2008ga,Machleidt:2011zz,Epelbaum:2019kcf},
the most extensively studied and best understood hadronic
interactions.
  As discussed in the Supplemental Material~\cite{SuppMat},
  such ambiguous contributions depend on several arbitrary parameters,
  whose values can be changed by
applying suitable UTs (subject to
the naturalness constraint).
Importantly, such UTs also induce
many-body interactions and exchange currents.
This shows, once again, that various scheme-dependent quantities  such as two-body interactions,
three-body forces (3BFs) and exchange currents must be used {\it consistently with each other}
to avoid any uncontrollable model dependence. 

\begin{figure}[t]
    \centering
    \includegraphics[width=0.99\linewidth]{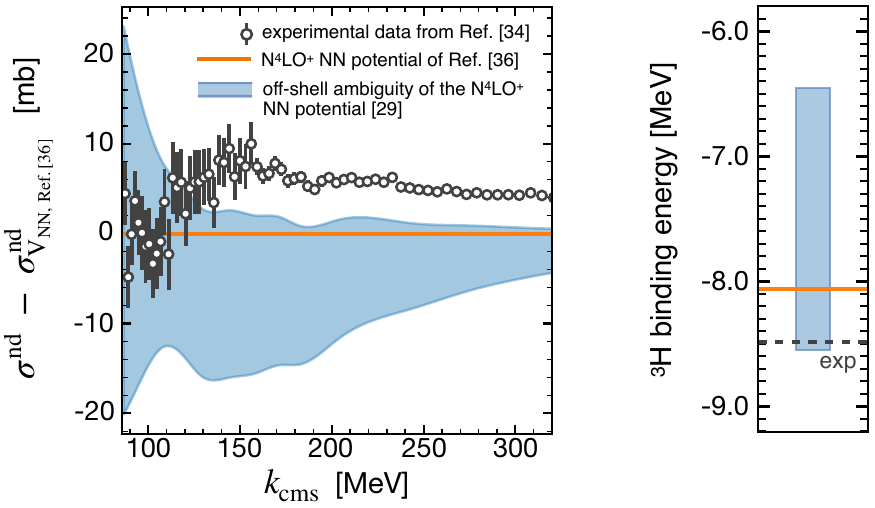}
    \caption{(Color online). Neutron-deuteron total cross section (left) and the $^3$H binding
      energy (right) calculated using the N$^4$LO$^+$ NN potential of Ref.~\cite{Reinert:2020mcu}
      (orange lines). Light-shaded blue bands show the results from the phase-equivalent but
      off-shell differing N$^4$LO$^+$ NN potentials as explained in the text. 
    }
    \label{fig4}
\end{figure}

The above power-counting-based arguments point towards a mild scheme-dependence of NN interactions
in chiral EFT. However, the situation is different for 3BFs, which provide small but important
corrections to the dominant pairwise forces and remain a challenging frontier in nuclear
physics \cite{Endo:2024cbz}, see also Refs.~\cite{ALICE:2022boj,ALICE:2023bny,Kievsky:2023maf}
for recent attempts to explore 3BFs through femtoscopy. 3BFs first appear at third order (N$^2$LO)
in chiral EFT, and thus are much more sensitive to the above mentioned off-shell ambiguities.
To illustrate this point we have generated a set of
NN N$^4$LO$^+$ potentials corresponding
to different choices of the off-shell short-range interactions, see~\cite{SuppMat} for details.
These potentials are nearly phase-equivalent in the two-body sector and equally valid from the EFT
point of view. However, they lead to vastly different predictions for three-nucleon observables
as visualized in Fig.~\ref{fig4}, which illustrates the well-known
inherent scheme dependence of
\phantom{\cite{Abfalterer:1998zz}} \hspace{-0.8cm}
3BFs \cite{Polyzou:1990hks}. In particular, the required 3BF contributions to the $^3$H
binding energy can be both attractive and repulsive depending on the off-shell behavior
of the employed NN interaction.  

{\it Discussion and conclusions.---}Hadronic correlations measured in ultra-relativistic
collisions are sensitive to the strong interactions. However, probing final state interactions
by means of the Koonin-Pratt formula violates basic principles of quantum mechanics if the source
model is regarded to be universal. Using an example of two strongly interacting distinguishable
particles, we have shown that off-shell ambiguities in the interaction can then translate into
a significant model dependence for $C({\bf k})$. It is important to emphasize that the problematic
universality assumption is essential as its relaxation sacrifices the predictive power of the
femtoscopy approach. 

The sensitivity of $C({\bf k})$ to the off-shell behavior of the strong force
decreases for source radii $r_0$ being large compared to the
interaction range 
since the large-distance behavior of the wave function $\Psi ({\bf r} , \, {\bf
  k})$ is unambiguously determined by the S-matrix \cite{Gmitro:1986ay,Lednicky:2005tb,Lisa:2005dd}.  
The interpretation of  $C({\bf k})$ in
terms of the average $| \Psi ({\bf r} , \, {\bf
  k}) |^2$
then becomes no more inherently problematic.
Large $r_0$-values, however, also reduce the strength of femtoscopic signals. The source size
in our example is, in fact, comparable to or even larger than those typically used in the literature
\cite{Fabbietti:2020bfg,ALICE:2020mfd,ALICE:2020ibs,ALICE:2022wpn,ALICE:2023bny}. 
Even smaller values for $r_0$ were obtained recently using a precise pion-kaon interaction \cite{Albaladejo:2025lhn}.

Finally, we also discussed off-shell ambiguities of nuclear interactions in chiral EFT. While
scheme-dependent contributions to the NN force are suppressed by power counting, the 3BF is
highly sensitive to the off-shell components of the NN potentials. The lack of the off-shell
consistency in femtoscopy analyses, therefore, raises concerns about the feasibility of
extracting 3BFs from heavy-ion collisions as anticipated in
Refs.~\cite{ALICE:2022boj,ALICE:2023bny,Kievsky:2023maf}. 

  The shortcomings of the femtoscopy approach due to the
  intrinsic scheme dependence of
  hadronic interactions, as discussed in
  our paper, must be 
taken into account in the uncertainty analysis of heavy-ion
experiments, especially when it comes to precision 
studies. Future research should also explore and
quantify the reduction of the model dependence in the calculated
correlation functions by employing specific constraints on the
interaction models (like, e.g., pion dominance at  large distances).  

\begin{acknowledgments}
One of the authors (EE)  is grateful to the organizers of the
``Three-body femtoscopy meeting'' in Prague, March 4-6, 2025, for
their hospitality and appreciates useful discussions with Raffaele Del
Grande
and Alejandro Kievsky.  This work has been supported by the European
Research Council (ERC) under the European Union’s Horizon 2020
research and innovation programme (grant agreement No.~885150 and
No.~101018170), by the MKW NRW under the funding code NW21-024-A, by
JST ERATO (Grant No. JPMJER2304), by JSPS KAKENHI (Grant
No. JP20H05636), by the Deutsche Forschungsgemeinschaft (DFG, German Research Foundation) under Germany's
Excellence Strategy – EXC 3107 – Project-ID~533766364, and by the Chinese Academy of Sciences (CAS) President’s International Fellowship Initiative (PIFI) (Grant No. 2025PD0022).
\end{acknowledgments}


\beginsupplement
\onecolumngrid


\newpage

\setcounter{page}{1}

\section{Supplemental Material}

In Sec.~I of this Supplemental Material, we extend the arguments given
in the
main text to proton-proton correlation functions. We also demonstrate that off-shell ambiguities of the
strong proton-proton interaction potentials significantly affect the
extracted radius of the source term if the latter is assumed to have a
universal (Gaussian) form. Finally, Sec.~II of the Supplemental Material
provides details on the construction of the phase-equivalent
two-nucleon potentials in chiral EFT discussed in the main text.

\subsection{I.~Femtoscopy analysis of proton-proton correlation functions}

In this section we repeat the Gedankenexperiment of the main
text, considered there for the case of distinguishable spinless particles in the absence of the
Coulomb interaction, for the two-proton system. The reference
strong-interaction potential (Alice) is chosen to be the proton-proton
N$^4$LO$^+$ chiral EFT potential of Ref.~\cite{Reinert:2017usi} with the cutoff
parameter $\Lambda = 450$~MeV. 

\subsubsection{A.~Gedankenexperiment for the two-proton system}

To calculate the proton-proton correlation function we use the
Koonin-Pratt formula in Eq.~(\ref{KooninPratt}). The coordinate-space
proton-proton scattering wave function $\Psi (r ,\, k) $ can be straightforwardly
obtained using the Vincent-Phatak method 
\cite{Vincent:1974} in the partial-wave basis. 
In the sum over the partial waves, we truncate the strongly
interacting channels at a maximum total angular momentum
$j=j_{\mathrm{max, int}}$ and supplement the sum with non-interacting
(free Coulomb) wave functions up to $j=j_{\mathrm{max, free}}$: 
\begin{equation}
  \label{WFpp}
|\Psi (r ,\, k) |^2 = \sum_{j=0}^{j_{\mathrm{max, int}}}
\frac{2j+1}{2}  \, \sum_{s,l',l} \big|\Psi^{j}_{s,l',l} (r ,\, k)\big|^2 
+ \sum_{j=j_{\mathrm{max, int}}}^{j_{\mathrm{max, free}}}
\frac{2j+1}{2} \,  \sum_{s,l'} \bigg|\frac{F_{l'}(\eta, rk)}{rk}\bigg|^2, 
\end{equation}
where $\Psi^{j}_{s,l',l} (r ,\, k)$ is the proton-proton scattering
wave function in the partial-wave basis,
$l$ and $l'$ are the orbital angular momentum quantum numbers while
$s=0, 1$ is the total spin. Further, $F_{l'}(\eta, rk)$ denotes  
the regular Coulomb wave function, $\eta=m_p\alpha_C/(2k)$ is  the
Sommerfeld factor,  $\alpha_C =e^2/4\pi \simeq 1/137$ refers to the fine-structure constant, while 
$k$ and $m_p$ are the on-shell scattering momentum and the proton
mass, respectively. 
The overall prefactor of $1/2$ in both terms on the
right-hand side of Eq.~(\ref{WFpp}) originates from the antisymmetrization  
factor of $2$ and the spin-averaging factor $ 1/[(2s_1 + 1)(2s_2 + 1)] = 1/4$.
In the actual calculations presented below, we set $j_{\mathrm{max, int}}=2$ and $j_{\mathrm{max, free}}=25$.

To calculate the scattering wave function  $\Psi^{j}_{s,l',l} (r ,\,
k)$ using the Vincent-Phatak method we first introduce the screened
Coulomb potential defined as $V_C^R (r) = \alpha_C/r$ for $r \leq R$ and $V_C^R (r)
= 0$ for
$r > R$. The screening radius $R$ must be chosen sufficiently large
compared to the range of the strong interaction. On the other hand,
too large values of $R$ lead to numerical instabilities due to the
singular behavior of the Coulomb potential at zero momentum transfer.  
In our numerical implementation, we set $R = 12$~fm.
We have explicitly verified, that the wave function has already reached its 
asymptotic behavior at the distances of $11\,\mathrm{fm} \leq r \leq
12\,\mathrm{fm}$. 
We solve the Lippmann-Schwinger equation for the 
screened Coulomb plus strong interaction potentials in the
momentum-space partial-wave basis to obtain the T-matrix
$T^{j}_{s,l',l}$. The proton-proton scattering wave function in
coordinate space can then be written in the uncoupled case as
\begin{equation}
  \label{WFVP}
\Psi^{j}_{s,l',l} (r ,\, k) = \begin{cases}
  A \big[\delta_{l'l} \, j_{l'}(kr)+
  2 m_p \int_{0}^{\infty} dp p^2 \, j_{l'}(pr) 
\,   T^{j}_{s,l',l} (p, k)/(k^2-p^2+i\epsilon) \big], & \text{for}\; r \leq R\,,\\[10pt]
  F_{l'}(\eta, rk)/(rk)+\frac{1}{2i}\big[G_{l'}(\eta, rk)/(rk)+i F_{l'}(\eta, rk)/(rk)\big]
  (S-1), & \text{for}\;r > R\,.
\end{cases}
\end{equation}
Here, $G_{l'}(\eta, rk)$ denotes the irregular Coulomb wave function,  
$j_{l'}(pr)$ is the spherical Bessel function, and $S = e^{2i\delta (k)}$ is the  
S-matrix of the strong plus Coulomb interaction relative to the Coulomb asymptotic
states. The corresponding phase shift $\delta (k)$ is determined by
matching the logarithmic derivatives of the wave function in
Eq.~(\ref{WFVP}) at $r = R$. The normalization constant $A$ is subsequently
determined by matching the wave functions at  $r = R$.
The extension to coupled channels can be straightforwardly achieved by replacing
the S- and T-matrices and the functions $F_{l'}(\eta, rk)$ and
$G_{l'}(\eta, rk)$ by the corresponding $2 \times 2$ matrices as
described in appendix C of Ref.~\cite{EPELBAUM2005362}.  

To compute the correlation function obtained by Bob one can proceed
in two different but equivalent ways by either applying the UT directly to the scattering wave function or,
alternatively, by transforming the interaction potential as done in
the main text. More precisely, the
unitarily transformed strong proton-proton potential used by Bob is defined by 
\begin{eqnarray}
\label{UT_potential_pp}
\hat V_{\rm Bob} &=& \hat U \bigg( \frac{\hat p^2}{2\mu} + \hat V_{\rm
                     Alice} + \hat V^R_{\rm
                     C} \bigg) \hat U^\dagger
- \frac{\hat p^2}{2\mu}  - \hat V^R_{\rm
                     C}\,.
\end{eqnarray}
The usage of the screened instead of the full Coulomb potential in the
above equation is justified for the screening radius chosen much
larger than the range of the unitary transformation. For the
transformation given in Eq.~(\ref{SauerUnitarytrafo}), this condition
corresponds  to $R \gg \alpha^{-1}$.
\begin{figure}[t]
    \centering
    \includegraphics[width=0.8\linewidth]{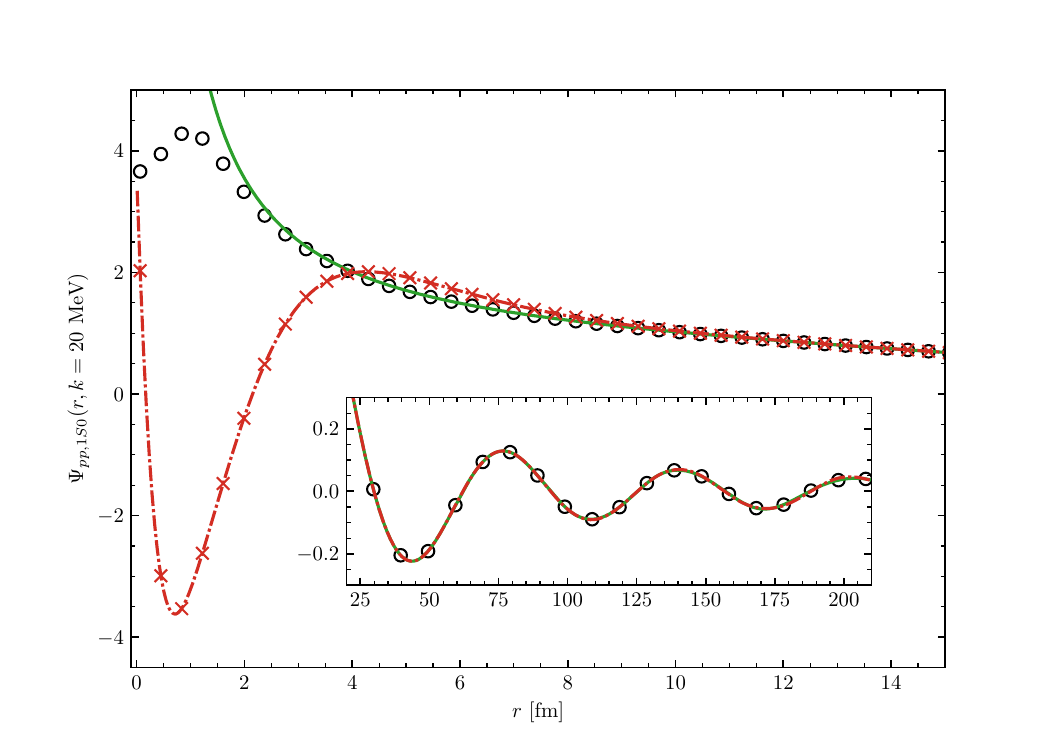}
    \caption{
      (Color online). The $^1$S$_0$
      proton-proton scattering wave function for $k=20$~MeV. The
      results of Alice are shown by open black circles. Red
      dash-dotted lines show the wave function obtained by applying
      the UT of Eq.~(\ref{SauerUnitarytrafo}) with the set-I parameters directly on the
      wave function of Alice.
      Red crosses show the wave function calculated from the potential
      $V_{\rm Bob\mbox{-}I}$ obtained using
      Eq.~(\ref{UT_potential_pp}).  Green solid lines show the
      asymptotic behavior of the wave function in terms of Coulomb
      wave functions and the phaseshift.
    } 
    \label{pp_WF_withUT}
\end{figure}
To demonstrate numerical stability
of our calculations, we show in Fig.~\ref{pp_WF_withUT} for one choice
of parameters of the UT (set-I) that the two different ways of computing the
transformed wave function indeed lead to the same results.  
As expected, the unitarily transformed (Bob-I) and original (Alice)
wave functions differ substantially in the interior region but coincide at large distances.
Therefore, extracting the phase shift from the asymptotic (large-distance) behavior  
of the wave function yields identical results, regardless of whether or not the  
UT is applied.

\begin{figure}[t]
    \centering
    \includegraphics[width=0.7\linewidth]{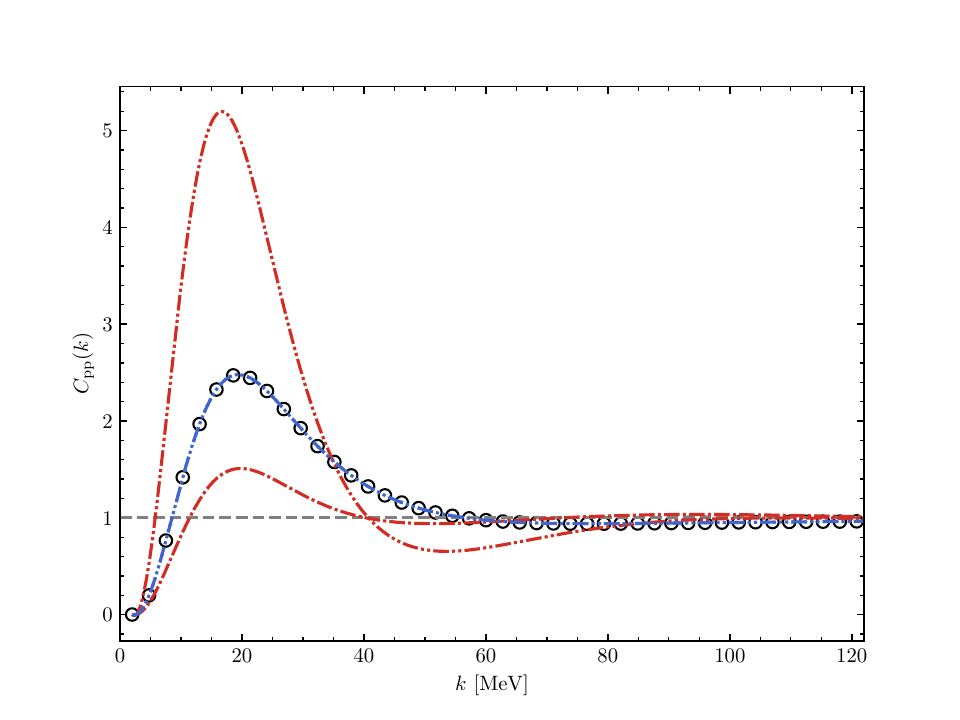}
    \caption{
      (Color online). Proton-proton  correlation functions calculated by Alice and Bob. The results of
      Alice are shown by black open circles. Red dash-dotted and dash-double-dotted lines show the
      correlation functions calculated by Bob using the interactions  $V_{\rm Bob\mbox{-}I}$ and
      $V_{\rm Bob\mbox{-}II}$, respectively, and the source term $S_{12}^{\rm Alice}$ (assumed to be universal).
      The overlapping blue dotted and dashed lines show the results of Bob using the
      properly transformed source terms $S_{12}^{\rm Bob\mbox{-}I}$ and $S_{12}^{\rm Bob\mbox{-}II}$, respectively. 
      A source radius of $2$~fm was used here.
    }
    \label{Cpp_Bob_andAlice_R02fm}
\end{figure}

With the wave functions obtained as described above, it is now possible to compute  
correlation functions using the Koonin-Pratt formula. In  
Fig.~\ref{Cpp_Bob_andAlice_R02fm}, we compare the proton-proton correlation functions computed  
by Alice and Bob.
As expected, if the source term is assumed to be universal and is not consistently  
transformed alongside the wave function, significant differences in
the resulting correlation functions emerge, see the open circles and
red lines. This occurs despite the fact that all  
wave functions exhibit identical long-range behavior (as shown in the inset of Fig.~\ref{pp_WF_withUT}). 
These discrepancies highlight  
the importance of consistently treating both the wave function and the source within  
the same unitary framework. In the same figure, we also show by blue
lines the correlation function obtained by unitarily 
transforming the wave function and simultaneously transforming the
source term (thereby making the source term non-local). The resulting
correlation functions are in perfect agreement with the one 
calculated by Alice reflecting the fact that UTs do not change observable quantities.

\subsubsection{B.~Extraction of the source radius from experimental data}

Finally, we apply phase-equivalent potentials used by Alice and Bob
to analyze the experimental data on proton-proton correlations 
measured in $p + {\rm Nb}$ collisions by the HADES Collaboration
\cite{Adamczewski-Musch:2016} with the goal to extract the source
radius $r_0$. To this aim, we perform a total of four fits of the source radius $r_0$, each using a different  
potential, while always assuming a universal source term, i.e.~we evaluate  
the Koonin-Pratt formula in Eq.~(\ref{KooninPratt}) using the same source function,  
given in Eq.~(\ref{SourceAlice}), in all cases. The four potentials differ only by  
short-range UTs and thus produce identical scattering observables.
However, the short-distance behavior of the corresponding wave
functions is significantly different. The potentials used are $V_{\rm Alice}$, $V_{\rm Bob\mbox{-}I}$,  
$V_{\rm Bob\mbox{-}II}$ (as discussed in the main text), and an additional potential  
$V_{\rm Bob\mbox{-}III}$, where the parameters of the UT are chosen to
be $\alpha = 1.15\,\mathrm{fm}^{-1}$ and $\beta =
0.5\,\mathrm{fm}^{-1}$ (set-III). 

\begin{figure}[h]
    \centering
    \includegraphics[width=0.7\linewidth]{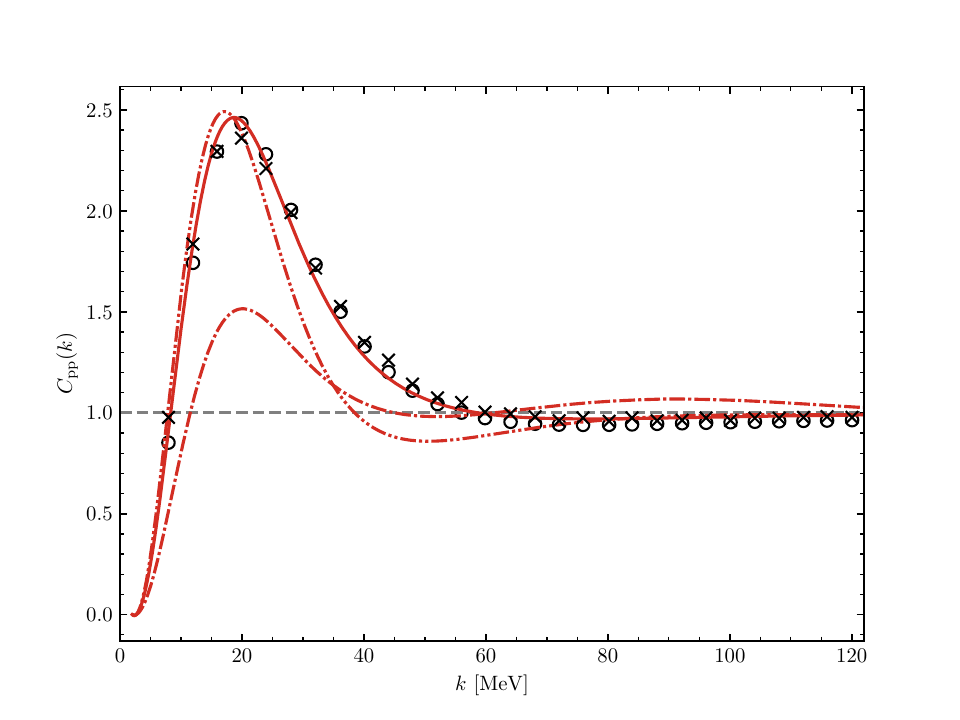}
    \caption{
      (Color online). Proton-proton correlation functions calculated by Alice and Bob. The results of
      Alice are shown by black open circles. Red dash-dotted, dash-double-dotted and solid lines show the
      correlation functions calculated by Bob using the interactions $V_{\rm Bob\mbox{-}I}$,
      $V_{\rm Bob\mbox{-}II}$, $V_{\rm Bob\mbox{-}III}$
      respectively. In  all cases, the source function $S_{12}^{\rm Alice}$ is used (assumed to be universal).
      The source radius $r_0$ is tuned for each individual interaction
      to obtain a best-possible agreement with the experimental data from Ref.~\cite{Adamczewski-Musch:2016} (black crosses).
      The fit results for the radius are $r_0^{\rm Alice}=2.03$~fm, $r_0^{\rm Bob\mbox{-}I}=1.75$~fm,  $r_0^{\rm
        Bob\mbox{-}II}=3.31$~fm and $r_0^{\rm Bob\mbox{-}III}=3.02$~fm.
      }
    \label{fittedCpp}
  \end{figure}
  
Fig.~\ref{fittedCpp} shows the  
four resulting correlation functions alongside the experimental data
from Ref.~\cite{Adamczewski-Musch:2016}. In each case, the source   
radius $r_0$ was treated as a free parameter and fitted to the data under the  
assumption of a universal source term.
The extracted values of $r_0$ vary significantly, ranging from  
$r_0 = 1.75\,\mathrm{fm}$ to $r_0 = 3.31\,\mathrm{fm}$. The correlation  
functions obtained from the original chiral EFT potential (Alice)
provide a good description of the experimental data using the source radius
of $r_0 = 2.03$~fm. On the other hand, the correlation functions
calculated from the potentials $V_{\rm
  Bob\mbox{-}I}$ and $V_{\rm  Bob\mbox{-}II}$ clearly fail to describe the
experimental data regardless of the employed $r_0$-value. One thus may
be tempting to conclude that a comparison of the calculated and
measured correlation functions provides a tool to uniquely 
determine the interaction compatible with the assumed local Gaussian
form of the source.  
This interpretation is, however, not correct as visualized by the
solid line in Fig.~\ref{fittedCpp}, which shows the correlation
function obtained using the $V_{\rm  Bob\mbox{-}III}$ potential. It
leads to a good description of the experimental data comparable to
that found by Alice yielding, however, a rather different value of the
source radius of $r_0 = 3.02$~fm.  

The above results demonstrate that the assumption of a universal source function can lead  
to significant ambiguities in the extracted form of the source term, even when using phase equivalent  
potentials that yield identical scattering observables. This highlights the need for a  
consistent treatment of the interaction together with the source term in femtoscopy analyses.

\subsection{II. Scheme dependence in nuclear chiral EFT and phase equivalent NN potentials at N$^4$LO$^+$}

In this Supplemental Material we focus on the off-shell ambiguities in the chiral nuclear interactions and describe the construction of the phase-equivalent but off-shell differing N$^4$LO$^+$ NN potentials. 

As emphasized in the main text, scheme dependence is an inherent
feature of nuclear interactions. Already the starting point for the
derivation of the nuclear interactions, the effective chiral Lagrangian
truncated at a given order in the derivative expansion, features a
certain degree of ambiguity as reflected in the freedom to perform
field redefinitions and apply equations of motion to eliminate various
terms. The derivation of nuclear potentials requires
off-the-energy-shell extensions of the (few-nucleon-irreducible) parts
of the scattering amplitudes and introduces additional scheme
dependence. For a description of various methods to derive
interactions in chiral EFT see Ref.~\cite{Epelbaum:2019kcf} and
references therein. A novel approach using the path integral
formalism, which is capable of dealing with regularized (non-local)
effective Lagrangians \cite{Krebs:2023gge}, has been proposed in
Ref.~\cite{Krebs:2023ljo}.  

Here, we focus on the expressions for the NN potentials, 3NFs and current operators given in Refs.~\cite{Epelbaum:1998ka,Epelbaum:1999dj,Epelbaum:2002gb,Epelbaum:2005fd,Epelbaum:2005bjv,Epelbaum:2007us,Bernard:2007sp,Bernard:2011zr,Krebs:2012yv,Krebs:2013kha,Kolling:2009iq,Kolling:2011mt,Krebs:2016rqz,
  Krebs:2019aka,Krebs:2020plh,Krebs:2020pii}, which were obtained
using the method of unitary transformations (UTs). In this approach,
one performes a UT of the effective pion-nucleon Hamiltonian to
decouple the states on the Fock space involving pions from the purely
nucleonic ones. This is achieved in a perturbative framework by
utilizing the standard chiral expansion \cite{Epelbaum:1998ka,
  Epelbaum:2007us}. To arrive at renormalized
expressions for the 3BFs and current operators it was necessary,
starting from N$^3$LO, to
systematically exploit the unitary ambiguity of the nuclear
interactions by considering a broad class of transformations on the
nucleonic subspace of the Fock space starting from N$^3$LO
\cite{Epelbaum:2005bjv,Bernard:2007sp,Krebs:2012yv,Kolling:2011mt,Krebs:2016rqz,Krebs:2019aka,Krebs:2020plh,
  Krebs:2020pii}. The corresponding unitary phases were then chosen in
such a way  that the resulting potentials remain finite after
renormalization (using dimensional regularization). For the considered
class of UTs, the resulting static expressions for the nuclear forces
at N$^3$LO and N$^4$LO turn out to be determined unambiguously, while
the leading relativistic corrections depend on two arbitrary real
parameters $\bar \beta_8$, $\bar \beta_9$ corresponding to the UTs
given in Ref.~\cite{Kolling:2011mt}.
In addition to the UTs
parametrized by the phases $\bar \beta_8$, $\bar \beta_9$,
nuclear
potentials at N$^3$LO feature the short-range ambiguity related to the UTs 
\begin{equation}
  \label{UTcont}
\hat U_{\rm short-range} = e^{- \gamma_1 \hat T_1 - \gamma_2 \hat T_2 - \gamma_3 \hat T_3}\,,
\end{equation}
where $\gamma_i$ are arbitrary real parameters and the short-range
anti-Hermitian NN operators $\hat T_i$ are given by
\cite{Reinert:2017usi}
\begin{eqnarray}
  \label{UTcontactGen}
  \langle {\bf p}_1' {\bf p}_2' | \hat T_1 |  {\bf p}_1 {\bf p}_2 \rangle
  &=& \frac{m}{8 \Lambda_{\rm b}^4} \big( p_1'^2 + p_2'^2 - p_1^2 -
p_2^2 \big) \,, \nonumber \\
  \langle {\bf p}_1' {\bf p}_2' | \hat T_2 |  {\bf p}_1 {\bf p}_2 \rangle &=& \frac{m}{8 \Lambda_{\rm b}^4} \big( p_1'^2 + p_2'^2 - p_1^2 -
p_2^2 \big) {\bm \sigma}_1 \cdot  {\bm \sigma}_2 \,, \nonumber \\
  \langle {\bf p}_1' {\bf p}_2' | \hat T_3 |  {\bf p}_1 {\bf p}_2 \rangle &=& \frac{m}{16 \Lambda_{\rm b}^4} \Big(  {\bm \sigma}_1 \cdot (
{\bf p}_1 - {\bf  p}_2 + {\bf  p}_1' - {\bf  p}_2') \;  {\bm \sigma}_2 \cdot (
{\bf  p}_1' - {\bf  p}_2' - {\bf  p}_1 + {\bf  p}_2 ) \nonumber \\
&&{} + 
{\bm \sigma}_1 \cdot ({\bf  p}_1' - {\bf  p}_2' - {\bf  p}_1 + {\bf  p}_2 )
\; {\bm \sigma}_2 \cdot ( {\bf  p}_1 - {\bf  p}_2 + {\bf  p}_1' - 
{\bf p}_2' 
) \Big) \,.
\end{eqnarray}
Here, ${\bf  p}_i$ and  ${\bf  p}_i'$ denote the incoming and outgoing momenta of the nucleon $i$ while 
$\Lambda_{\rm b}$ is the breakdown scale of chiral EFT, respectively.
Note that in line with the commonly used notation, we
do not show in Eq.~(\ref{UTcontactGen}) the overall factor of 
$(2 \pi)^3$ and the $\delta$-function corresponding to the total
momentum conservation. We further emphasize that we do not consider
here short-ranged UTs, whose  generators vanish in the NN cms
\cite{Girlanda:2020pqn}. 
The induced contributions to the nuclear forces have the form 
\begin{equation}
  \label{InducedContact}
\delta \hat V_{\rm short-range} = \Big[ \big(\hat H_{\rm kin} +  \hat V_{\rm nucl}^{\rm LO} \big), \; \big( \gamma_1 \hat T_1 + \gamma_2 \hat T_2 + \gamma_3 \hat T_3 \big) \Big] + \ldots \,,
\end{equation}
  where the ellipses refer to terms nonlinear in
  $\gamma_i$, which are suppressed by power counting.
In the two-nucleon system, the commutator with $\hat V_{\rm nucl}^{\rm LO}$ leads to short-range NN operators that can be absorbed into a redefinition of the contact interactions. On the other hand,  the commutator with $\hat H_{\rm kin}$ generates purely off-shell short-range interactions in the NN $^1$S$_0$,  $^3$S$_1$ and $^3$S$_1$- $^3$D$_1 $ partial waves. Up to N$^4$LO, the contact interactions in these channels have the form 
\begin{eqnarray}
\langle ^1{\rm S}_0, \, p' | \hat V_{\rm cont} | ^1{\rm S}_0, \, p \rangle  &=& \tilde
C_{^1{\rm S}_0} + C_{^1{\rm S}_0} (p^2 + p'^2) + D_{^1{\rm S}_0} p^2 p'^2 + D_{^1{\rm S}_0}^{\rm off} (p^2 -
                                                                           p'^2 )^2\,, \nonumber \\
 \langle ^3{\rm S}_1, \, p' | \hat V_{\rm cont} | ^3{\rm S}_1, \, p \rangle  &=& \tilde
C_{^3{\rm S}_1} + C_{^3{\rm S}_1} (p^2 + p'^2) + D_{^3{\rm S}_1} p^2 p'^2 + D_{^3{\rm S}_1}^{\rm off} (p^2 -
                                                                           p'^2 )^2\,, \nonumber \\ 
\langle ^3S_1 , \, p' | \hat V_{\rm cont} | ^3D_1 , \, p \rangle  &=&
C_{\epsilon_1} p^2 + D_{\epsilon_1} p^2 p'^2 + D_{\epsilon_1}^{\rm off} p^2(p'^2 - p^2)
\,, 
\end{eqnarray}
where $\tilde C_{i}$ are the LO low-energy constants (LECs), $C_{i}$ are the next-to-leading order (NLO) LECs while
$D_{i}$ and $D_{i}^{\rm off}$ are the N$^3$LO LECs. Notice that the
contact interactions proportional to $D_{^1{\rm S}_0}^{\rm off}$,
$D_{^3{\rm S}_1}^{\rm off}$ and $D_{\epsilon_1}^{\rm off}$ vanish in
the on-shell kinematics with $p' = p = k$ and thus cannot be
determined reliably from fits to NN scattering data. This can also be
understood from a different perspective by observing that the
considered short-ranged UTs generate the NN contact interactions whose
form is identical to those proportional to $D_{i}^{\rm off}$ when the
commutator in Eq.~(\ref{InducedContact}) is evaluated with $\hat
H_{\rm kin}$. This implies that the contact terms proportional to the
LECs  $D_{i}^{\rm off}$ can be rotated away by means of a suitably
chosen UT. Accordingly, the NN potentials of
Refs.~\cite{Reinert:2017usi,Reinert:2020mcu} were constructed using
the scheme  with  
\begin{equation}
  \label{convention}
D_{^1{\rm S}_0}^{\rm off} = D_{^3{\rm S}_1}^{\rm off} = D_{\epsilon_1}^{\rm off} = 0\,.
\end{equation}
It is important to emphasize that the considered UTs also induce 3BFs
when evaluating the commutator in Eq.~(\ref{InducedContact}) with $\hat
V_{\rm nucl}^{\rm LO}$ \cite{Reinert:2017usi,Girlanda:2023znc}. This
demonstrates, once again, that 3BFs can only be meaningfully defined
in conjunction with the NN potential, i.e., within a given scheme
fixed by a specific off-shell behavior of the NN interaction.   

The choice specified in Eq.~(\ref{convention}) and adopted in
Refs.~\cite{Reinert:2017usi,Reinert:2020mcu} is obviously just one
possibility out of infinitely many. From the EFT point of view, any values of $D_{i}^{\rm off}$ are acceptable provided the LECs of the NN contact interactions are of natural size.
To illustrate scheme dependence of the 3BF in chiral EFT, we have
generated a set of semi-local momentum-space regularized (SMS) NN
potentials \cite{Reinert:2017usi,Reinert:2020mcu,Epelbaum:2022cyo},
which are nearly phase equivalent but differ by the choices of the
off-shell LECs $D_{^1{\rm S}_0}^{\rm off}$, $D_{^3{\rm S}_1}^{\rm
  off}$ and $D_{\epsilon_1}^{\rm off}$. Specifically, we consider the
values 
\begin{equation}
  \label{off-shell_LECs}
D_{^1{\rm S}_0}^{\rm off} = D_{^3{\rm S}_1}^{\rm off}  = \pm 3,\,  0; \qquad  D_{\epsilon_1}^{\rm off} = \pm 1,\,  0
\end{equation}
in units of $10^4$~GeV$^{-6}$ used in Ref.~\cite{Reinert:2017usi} and employ the
cutoff value of $\Lambda = 450$~MeV. These units coincide with the
natural units of $4\pi/(F_\pi^2 \Lambda_{\rm  b}^4)$ used in 
Refs.~\cite{Epelbaum:2014efa,Epelbaum:2019kcf},
where $F_\pi$ and $\Lambda_{\rm b} $ denote the pion decay
constant and the breakdown scale of chiral EFT, respectively, if one
sets $\Lambda_{\rm b} \simeq 620$~MeV. 
To ensure approximate phase equivalence of the resulting potentials,
we stay at the highest available EFT order N$^4$LO$^+$, which provides
a sufficient flexibility to achieve a statistically perfect
description of NN data up through the pion production threshold, see
Refs.~\cite{Reinert:2017usi,Epelbaum:2022cyo} for details, and
restrict ourselves to the cutoff value of $\Lambda = 450$~MeV. 
The assumed range of values for $D_{^1{\rm S}_0}^{\rm off}$ and
$D_{^3{\rm S}_1}^{\rm off}$ can  be considered as fairly
conservative with regard to the naturalness assumption. For example,
with the off-shell LECs switched off, the resulting N$^3$LO LECs
$D_i$ fulfill $| D_i | \leq 2.7\times 10^4$~GeV$^{-6}$ for $\Lambda = 450$~MeV
\cite{Reinert:2017usi}.  For the LEC $D_{\epsilon_1}^{\rm off}$, the
values $D_{\epsilon_1}^{\rm off} = \pm 3$ turn out to cause
unrealistically strong modifications of the deuteron D-state
probability, which would render the resulting potentials impractical
for many-body applications \cite{Machleidt:2009bh}. We have therefore
limited ourselves to the values $|D_{\epsilon_1}^{\rm off}| \leq 1$.  

For each of the 26 combinations of the off-shell LECs in
Eq.~(\ref{off-shell_LECs}) beyond the already available one
corresponding to Eq.~(\ref{convention}), we have fitted the remaining LECs
accompanying the NN contact interactions to the  neutron-proton and
proton-proton scattering data up to $E_{\rm lab} = 280$~MeV, following
the same protocol and using the same database as employed in
Refs.~\cite{Reinert:2020mcu,Reinert:2022jpu}.
We emphasize that when changing the values of the off-shell LECs
and/or the LECs beyond the considered expansion order, the LECs
accompanying contact interactions at all orders must be refitted. This
procedure corresponds to implicit renormalization within
the considered EFT framework and 
ensures that all results are given in terms of physical
parameters (i.e., experimental data for observables used in the fit)
rather than bare LECs, see Ref.~\cite{Epelbaum:2019kcf} for details.  
For all considered cases, we found an essentially perfect description of the
neutron-proton (np) and proton-proton scattering data as reflected in the
$\chi^2/{\rm datum}$-values in the range of $\chi^2/{\rm datum} =
1.003 \ldots 1.007$. These results provide an important consistency
check of our calculations and show, in particular, that the
contributions in Eq.~(\ref{UTcontactGen}) that have been neglected at the
considered accuracy level based on the power-counting arguments are
indeed numerically negligibly small.

\begin{figure}[t]
    \centering
    \includegraphics[width=0.9\linewidth]{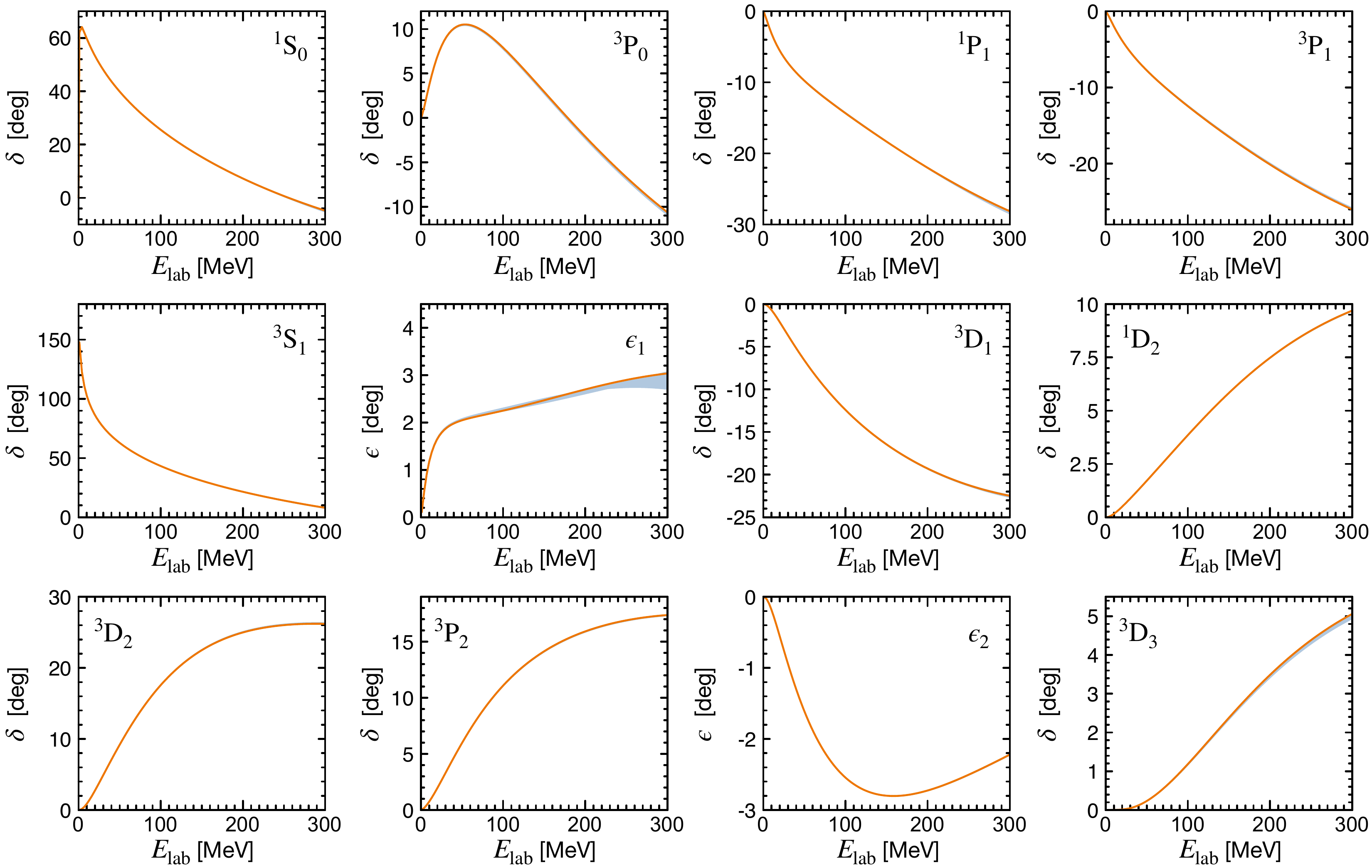}
    \caption{(Color online). The results for the neutron-proton phase
      shifts and mixing angles calculated using the 27 N$^4$LO$^+$
      potentials with different choices of the off-shell LECs
      specified in Eq.~(\ref{off-shell_LECs}) are shown by
      light-shaded blue bands. Phase shifts and mixing angles obtained
      using the potential from Ref.~\cite{Reinert:2020mcu} with $D_{^1{\rm S}_0}^{\rm off} = D_{^3{\rm
          S}_1}^{\rm off} = D_{\epsilon_1}^{\rm off} = 0$ are shown by
      orange lines. In all cases, the cutoff is set to $\Lambda = 450$~MeV.
    }
    \label{SM_fig1}
\end{figure}

In Fig.~\ref{SM_fig1}, we show the resulting np phase shifts and
mixing angles in low partial waves. These results demonstrate that the
27 potentials can indeed be regarded as essentially phase
equivalent. The visible (but tiny) dependence of the 
mixing angle $\epsilon_1$ on the off-shell LECs reflects the fact that
this particular observable is less well constrained by the
experimental data as compared to other phase shifts. For an
estimation of the EFT truncation uncertainty and other types of uncertainties see
Refs.~\cite{Epelbaum:2019kcf,Epelbaum:2022cyo,Reinert:2022jpu}.   

The predictions for the
deuteron properties for the 27 N$^4$LO$^+$ potentials are collected in Table \ref{SM_tab}. 
\begin{table}[tb]
\begin{tabular*}{\textwidth}{@{\extracolsep{\fill}}ll@{\extracolsep{\fill}}l@{\extracolsep{\fill}}r@{\extracolsep{1pt}}l@{\extracolsep{1pt}}r@{\extracolsep{1pt}}l@{\extracolsep{1pt}}l}
        \toprule
                 & \hspace{-0.75cm} N$^4$LO$^+$,  Ref.~\cite{Reinert:2020mcu} &
                                                               \hspace{-1.5cm} N$^4$LO$^+$,
                                                               off-shell
                                                               LECs
                                                               from Eq.~(\ref{off-shell_LECs})
  & \multicolumn{3}{c}{Empirical} \\
        \midrule
        $B_d$ [MeV]   \phantom{xxxxxxxxxxxxx}                          & $2.2246$$^\star$\phantom{xxxxxxxxxxxxxxx}  &   $2.2246$$^\star$&  $2.22456614$ & $(41)$ & \cite{Kessler:1999zz} \\
        $A_S$ [fm$^{-1/2}$]                    &   $0.8846$ &
                                                              $0.8845\ldots 0.8848$ &  $0.8845$ & $(8)$  & \cite{deSwart:1995ui} \\
        $\eta$                                 &   $0.0261$ &
                                                              $0.0260\ldots 0.0263 $ &    $0.0256$ & $(4)$  & \cite{Rodning:1990zz} \\
        $r_m$ [fm]                             &   $1.9662$   &
                                                                $1.9588\ldots 1.9709 $   &   --- &  & \\
        $Q_0$ [fm$^2$]                         &    $0.275$  &   \hspace{0.1682cm}$0.269  \ldots 0.280$   &  --- & & \\
        $P_D$ [\%]                             &     $4.79$ &    \hspace{0.318cm}$3.80 \ldots 6.33$    &        --- &        & \\
        \bottomrule
    \end{tabular*}
    \begin{tabular*}{\textwidth}{@{\extracolsep{\fill}}l}
        $^\star$The deuteron binding energy has been taken as input in the fit. 
    \end{tabular*}
       \caption{Deuteron binding energy $B_d$, asymptotic S-state
      normalization $A_S$, asymptotic D/S-state ratio $\eta$, matter
      radius $r_m$, leading contribution to the quadrupole moment
      $Q_0$ and D-state probability $P_D$ obtained using the SMS N$^4$LO$^+$
      potential of Ref.~\cite{Reinert:2020mcu} and the 27 N$^4$LO$^+$
      potentials with the off-shell LECs $D_i^{\rm
        off}$ set according to Eq.~(\ref{off-shell_LECs}) are given in
      the second and third columns, respectively. The cutoff is set to $\Lambda = 450$~MeV.
   }
    \label{SM_tab}
\end{table}
The resulting values for the S-state normalization observable $A_S$ and
the asymptotic D/S-state ratio $\eta$ show, similarly to the phase
shifts, a very small sensitivity to the off-shell LECs and agree with
the experimental data within errors, see Ref.~\cite{Epelbaum:2022cyo}
for the error analysis. On the other hand, the 
matter radius  $r_m$ (i.e., the expectation value of the relative
distance between the nucleons), the quadrupole moment 
of the deuteron wave function $Q_0$ and the D-state probability $P_D$
are well known not to correspond to observable quantities, see
e.g.~\cite{Friar:1979zz}.  Not surprisingly, our
predictions for these quantities demonstrate a significant degree of scheme
dependence.
In contrast to $r_m$ and $Q_0$, the deuteron charge radius and quadrupole
moment related to the electric and quadrupole form factors $G_C(q^2)$
and $G_Q (q^2)$, respectively,  via $r_c =
6 \frac{1}{G_C(0)} \frac{dG_C(q^2)}{dq^2}\big|_{q^2 = 0}$ and $Q =
\frac{1}{m_d^2} G_Q (0)$ with $m_d$ denoting the deuteron mass are, of
course, observables. The scheme-dependent quantities $r_m$ and $Q_0$ provide the bulk
contributions to the charge radius and the quadrupole moment as they
originate from the dominant single-nucleon charge density, but one
also needs to take into account two-body contributions to the charge
density operator, which are often referred to as ``meson-exchange currents''.
These (scheme-dependent) meson-exchange currents must be chosen consistently
with the nuclear interactions to ensure that the resulting form
factors are unambiguous. Indeed, it is easy to
see that the unitary transformation in Eq.~(\ref{UTcont}) generates
short-range meson-exchange currents when acting on the single-nucleon
charge density \cite{Filin:2020tcs}, rendering the calculated form factors independent of the
arbitrary phases $\gamma_i$. Note that the spread in the obtained results for
$r_m$ and $Q_0$ agrees qualitatively with the expected size of
meson-exchange contributions from the power-counting, see
Refs.~\cite{Filin:2020tcs} for details.   

\begin{figure}[t]
    \centering
    \includegraphics[width=0.8\linewidth]{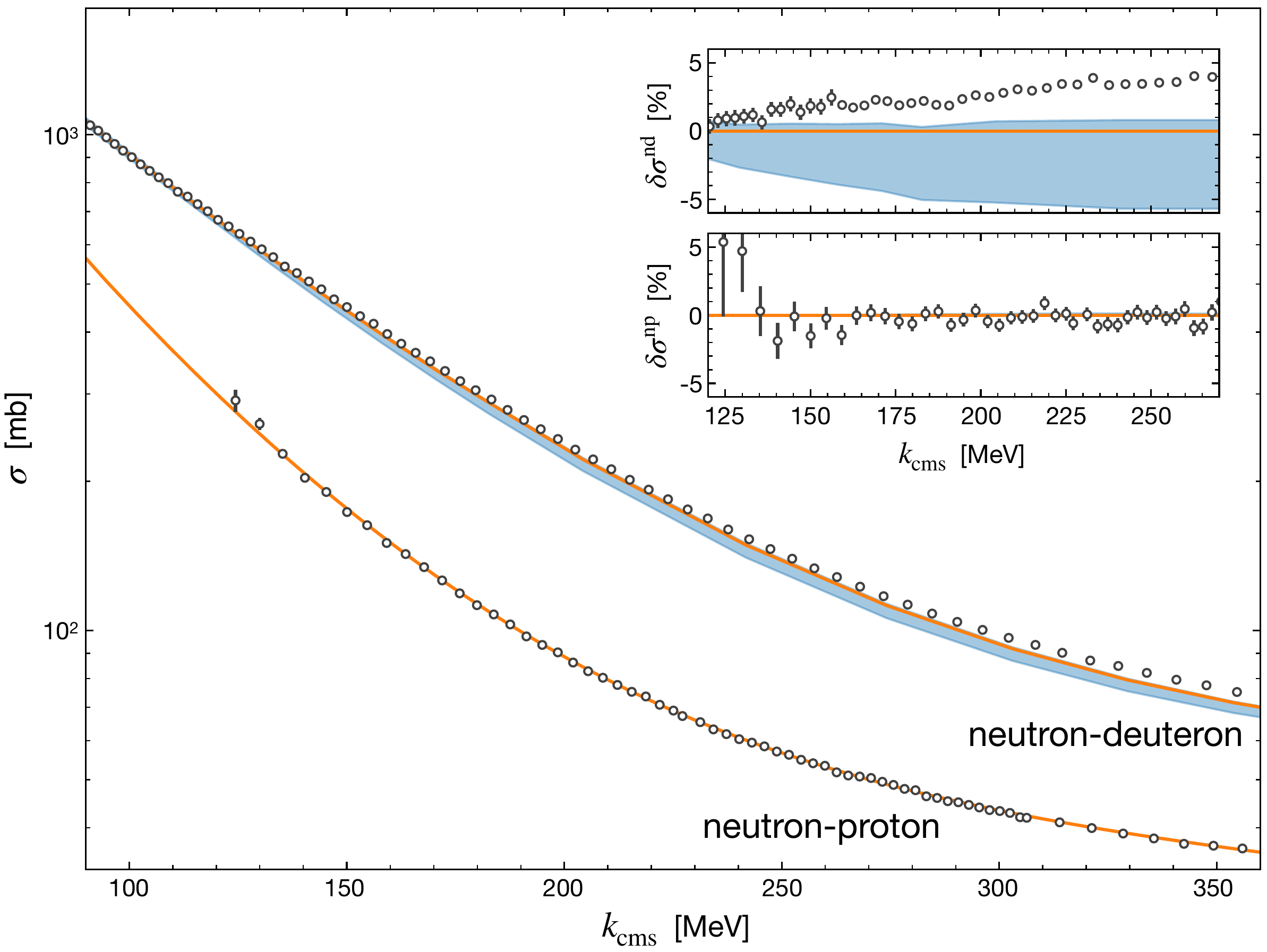}
    \caption{(Color online). Neutron-proton and
        neutron-deuteron total cross section as a function of the cms
        momentum $k_{\rm cms}$. The upper and lower inset plots show
        the relative cross section differences for nd
        and np
        scattering, respectively, defined as $\delta \sigma \equiv (
        \sigma -  \sigma_{\rm V_{NN, \, [36]}})/\sigma_{\rm V_{NN, \,
            [36]}}$. Light-shaded blue bands (not visible for
        neutron-proton scattering) are obtained from the 27
        nearly phase-equivalent N$^4$LO$^+$ NN potentials and
        illustrate inherent scheme dependence of nuclear interactions in chiral
        EFT. Experimental data for the nd and np total cross section
        are taken from Refs.~\cite{Abfalterer:1998zz} and \cite{Lisowski:1982rm}, respectively. 
    }
    \label{SM_fig2}
\end{figure}

Finally, in Fig.~\ref{SM_fig2}, we show the results for the
np and neutron-deuteron (nd) total cross section in comparison
with the experimental data. As already evident from the results for
phase shifts, the np cross section comes out nearly identical for all
27 potentials with the maximal relative
differences below two-permille level. On the other hand, the nd cross section calculated using
the NN forces only shows a significant scheme dependence, as already visible from Fig.~\ref{fig4}.  
The difference between the nd experimental data and calculations 
utilizing a particular NN potential gives the contribution of the 3NF
needed to reproduce the cross section data. The large spread in the
predictions for three-nucleon observables from different but nearly
phase equivalent N$^4$LO$^+$ NN potentials, reflected in the
light-shaded blue bands in Figs.~\ref{fig4} and \ref{SM_fig2},
demonstrates the strong inherent scheme dependence of 
the 3BF in the framework of chiral EFT. 

The results for the nd total cross section and the $^3$H
binding energy shown in Figs.~\ref{fig4} and \ref{SM_fig2} agree well
with estimations based on the power counting, see
Ref.~\cite{LENPIC:2015qsz} for a related discussion. For example,
given the expansion parameter in chiral EFT of the order of $Q \sim
1/3$ in chiral EFT and the expectation value of the NN interaction for
$^3$H of $\langle V_{\rm NN} \rangle \sim 40 \ldots 50$~MeV,  one may
expect the 3BF contribution to the  $^3$H binding energy roughly of
the order of  $\langle V_{\rm 3NF} \rangle \sim  Q^3 \, \langle V_{\rm NN} \rangle  \sim 1.5$~MeV, where we took into
account the first appearance of the 3BF at N$^2$LO ($Q^3$).  
Furthermore, the approximately constant deviations between the experimental data for
the nd total cross sections and the blue band comprising the results based on the
27 N$^4$LO$^+$ NN potentials in Fig.~\ref{fig4}  translate into the relative
deviation that grows with energy and reaches about $\sim 10 \ldots
20$\% at $E_{\rm lab} = 200$~MeV, which corresponds to $k_{\rm cms}
\sim 400$~MeV. This pattern and the amount of underprediction of
$\sigma^{\rm nd}$ are also well in line with the expected size of
the 3BF contributions based on the chiral power counting
\cite{LENPIC:2015qsz}. In particular, the growing contributions of the
3NF at increasing energies, also seen in the context of the nuclear
equation of state, are consistent with the assumed EFT
expansion parameter $Q \sim k_{\rm cms}/\Lambda_{\rm b}$.  
Last but not least, we emphasize that the leading 3BF at N$^2$LO was
found to increase the nd total cross section in Ref.~\cite{LENPIC:2022cyu}, thereby
improving the agreement between theory and experiment.   

\end{document}